                                                                  \documentclass[12pt,preprint]{aastex}

\shorttitle{The twist and helicity of a eruptive filament}
\shortauthors{Yan et al.}

\begin{document}

\title{Kink instability evidenced by analyzing the leg rotation of a filament}
\author{X. L. Yan\altaffilmark{1,2}, Z. K. Xue\altaffilmark{1}, J. H. Liu\altaffilmark{3},
L. Ma\altaffilmark{1}, D.F. Kong\altaffilmark{1}, Z.Q. Qu\altaffilmark{1}, Z. Li\altaffilmark{2}}

\altaffiltext{1}{Yunnan Observatories, Chinese Academy of
              Sciences, Kunming 650011, China.}
\altaffiltext{2}{Key Laboratory of Modern Astronomy and Astrophysics (Nanjing University), Ministry of Education, Nanjing 210093, China.}

\altaffiltext{3}{Department of Physics, Shijiazhuang University, Shijiazhuang 050035, China.}
\begin{abstract}

Kink instability is a possible mechanism for solar filament eruption. However, the twist of a solar filament is very difficult to directly measure from observation.
In this paper, we carried out the measurement of the twist of a solar filament by analyzing its leg rotation. An inverse S-shaped filament in active region NOAA 11485 was observed by the Atmospheric Imaging Assembly (AIA) of Solar Dynamics Observatory (SDO) on 2012 May 22. During its eruption, the leg of the filament exhibited a significant rotation motion. The 304 \AA\ images were used to uncurl along the circles, whose centers are the axis of the filament's leg. The result shows that the leg of the filament rotated up to about 510 degrees (about 2.83$\pi$) around the axis of the filament within twenty-three minutes. The maximal rotation speed reached 100 degrees per minute (about 379.9 km/s at radius 18$^\prime$$^\prime$), which is the fastest rotation speed that has been reported. We also calculated the decay index along the polarity inversion line in this active region and found that the decline of the overlying field with height is not so fast enough to trigger the
torus instability. According to the condition of kink instability, it is indicating that the kink instability is the trigger mechanism for the solar filament eruption.
\end{abstract}



\section{Introduction}
Understanding the equilibrium of magnetic loops in the corona is one of the fundamental problems in solar physics. The stability of current-carrying force-free magnetic field in the corona was studied by many authors (Gold \& Hoyle 1960; Sakurai 1976; Hood \& Priest 1979; Titov \& Demoulin 1999; Baty 2001; T\"{o}r\"{o}k, Kliem, \& Titov 2004; Kliem, Titov, \& T\"{o}r\"{o}k 2004). It was found that the stability of magnetic loops is mainly
controlled by the total twist, the winding of the magnetic field lines about the flux rope axis.
The twist of the flux rope is the angle through which a magnetic field line is turned in going from one end of the rope to the other. The total twist, $\Phi$, of the helical-field structure is written as:
\begin{equation}
\label{1}
\Phi=\frac{lB_\phi(r)}{rB_z(r)},
\end{equation}
where $l$ and $r$ are the length and the radius of the magnetic flux rope, respectively. $B_\phi(r)$ and $B_z(r)$ are the azimuthal and the axial components of the equilibrium magnetic field, respectively.
For the cylindrically symmetric flux rope, the instability of the solar flux rope occurs, when the twist exceeds a critical value, $\Phi_c$ (Vr\v{s}nak et al. 1988; Hood \& Priest 1979, 1981; Vr\v{s}nak et al. 1991).  For example, when $\Phi$ $>$ $\Phi_c$ $=$ 2$\pi$, the uniformly twisted force-free toroidal or periodic cylindrical configuration becomes kink-unstable. The instability threshold reaches 2.49$\pi$ for line-tying configuration (Hood \& Priest 1981). For the occurrence of the kink instability to a cylindrical flux rope, the following condition must be satisfied:
\begin{equation}
\label{2}
kR \leq 0.272\mu R, 
\end{equation}
where $\mu$ is the linear force-free factor and $k$ is the ratio of 2$\pi$ to the length of the flux rope. When $\mu$$R$ $\cong$ 2.405, the instability condition becomes $kR$ $\leq$ 0.654 (see Rust \& Kumar 1996), where $R$ is the radius of the cylindrical region.

T\"{o}r\"{o}k \& Kliem (2003) simulated the formation of a twisted magnetic flux rope, which is created from an initially potential coronal flux tube driven by slow photospheric vortex motion at two photospheric flux concentrations, using the compressible zero-beta ideal MHD equation. They found that there is a critical end-to-end twist in the range 2.5$\pi$ $<$ $\Phi_c$ $<$ 2.75$\pi$ for a particular set of parameters, describing the initial potential field. If the twist exceeds the value of $\Phi_c$, the flux rope cannot reach a new equilibrium in their simulation.

Rotating magnetic structures driven by underlying photospheric vortex flows were observed by several authors (Zhang et al. 2011; Wedemeyer-B\"{o}hm et al. 2012; Yan et al. 2013). Vr\v{s}nak (1980) observed the detwisting motions with radial velocity of 100 km/s in prominences by using shapes and profiles of the H$\alpha$ line. The rotation and non-radial motion of solar filaments were often observed during their eruptions (Ji et al. 2003; Green et al. 2007; Jiang et al. 2009; Bi et al. 2013). The prominence-related rotation phenomenon was named as giant tornadoes, which were recently concerned due to high-quality observational data of SDO (Li et al. 2012; Su et al. 2012; Wedemeyer et al. 2013; Panesar et al. 2013). However, the twist of the filament has not been accurately measured. Thanks to high temporal and spatial data of SDO, we can carry out the measurement of the twist of the filament via analyzing the filament's leg rotation during the sigmoid filament eruption on 2012 May 22.

\section{Observation and method}
\subsection{Observation}
The Atmospheric Imaging Assembly (AIA; Lemen et al. 2012) on board Solar Dynamics Observatory (SDO)
provides multiple simultaneous high resolution full-disk images of the transition region and the corona.
The spatial and temporal resolutions of AIA are 1.5$^\prime$$^\prime$ and 12 s, respectively. The observation of the AIA includes seven EUV, two UV, and one visible-light wavelength bands. The temperature of the EUV emission line covers a wide range
from 6 $\times$ 10$^4$ K to 2 $\times$ 10$^7$ K. The field of view of AIA can be extended to 1.3 solar radius. In this paper, we only used the 304 \AA\ (He II, T = 0.05 MK) images to trace the filament's leg rotation during the filament eruption. Full disk line-of-sight magnetograms at 45 s cadence with a precision of 10 G
observed by Helioseismic and Magnetic Imager (HMI; Schou et al. 2012) are used to extrapolate the potential field in the corona.
\subsection{Method}
First of all, we must find the rotation center of the right leg of the filament. Second, once the center is determined, the 304 \AA\ images are uncurled to a polar
coordinate r - $\theta$ frame from the initial Cartesian x - y frame (Brown et al. 2003; Hardersen et al. 2013). The circles around the axis of the filament leg are uncurled counterclockwise, starting from a northward pointing chord (the black line from the center to the north), as illustrated in Figure 1. Third, we chose the five circles around the rotation center. The difference between the radii of adjacent circles is 12$^\prime$$^\prime$ from the inside to the outside. The resolution used in the angular direction is 1$^{\circ}$ to make the time slices. From the time-slices taken from 304 \AA\ images at a constant radius, the rotation speed, rotation angle $\theta$, and time t would be calculated by tracing the features of the filament. Note that the rotation angle was measured from the features that the filament was projected on the plane coordinate.

\section{Result}
An inverse S-shaped filament indicated by the white arrow in Fig.1 in active region NOAA 11485 was observed by AIA on board SDO on 2012 May 22. Figure 2 shows the rotating process of the right leg of the filament acquired at 304 \AA\ from 01:34:08 UT to 02:17:20 UT on 2012 May 22. The white box in Fig. 2a denotes the position of the inverse S-shaped filament. The arrows in Figs. 2(b), 2(c), and 2(d) indicate the right leg of the filament. The left part of the filament first began to rise and rotated counterclockwise. Then the expansion of the filament was accompanied by the rotation of the filament. Following the rising and the rotation of the left part of the filament, the right part of the filament began to rotate around the axis of the filament. Finally, the filament attained a circular shape. The process of the filament eruption can be seen from the supplementary movie 1.

In order to show the filament eruption clearly, we also presented the 304 \AA\ running-difference images from 01:54:08 UT to 02:16:20 UT on 2012 May 22 in Fig. 3. The supplementary movie 2 shows the process of the filament eruption by using the 304 \AA\ running-difference images.

During the filament eruption, both the filament and its right leg exhibited a significant counterclockwise rotation. With the method described in section 2.2, the rotation of the leg of the filament was measured as horizontal movement at certain radius. Figure 4 gives four examples
showing the change of the rotation angle at different radius. The rotation angle can be
effectively extracted from time slices at radii 18$^\prime$$^\prime$, 30$^\prime$$^\prime$, 42$^\prime$$^\prime$, and 66$^\prime$$^\prime$, respectively.
The rotation of the black and the white features can be seen clearly from the time slices. From the paths of the features, the rotation angle of the features every one minute can be determined. The rotation angles at different radii were calculated along the red dashed lines in Fig. 4. After the rotation angles were obtained, the average rotation speeds (degrees per minute) can be calculated. Furthermore, the angular speed can be converted to a speed in km/s.



Figure 5 shows the angle, angle speed, and rotational speed at radii 18$^\prime$$^\prime$ (black line), 30$^\prime$$^\prime$ (blue line), 42$^\prime$$^\prime$ (green line), and 66$^\prime$$^\prime$ (red line) from 02:00 UT to 02:26 UT on May 22, 2012. We adopted the average values of the three repeated measurements of the angles along the red dashed lines in Fig. 4. The rotation angle reached 505 and 510 degrees (about 2.8$\pi$) at radii 18$^\prime$$^\prime$ and 30$^\prime$$^\prime$ within twenty-three minutes. The rotation angle is 350 degrees within eighteen minutes at radius 42$^\prime$$^\prime$ and the rotation angle is 160 degrees within twelve minutes at radius 66$^\prime$$^\prime$. The average rotation speeds are 21.9, 22.2, 13.6, and 13.3 degrees per minute for radii 18$^\prime$$^\prime$, 30$^\prime$$^\prime$, 42$^\prime$$^\prime$, and 66$^\prime$$^\prime$, respectively. The maximal rotation speed reached 100 degrees per minute (about 379.9 km/s at radius 18$^\prime$$^\prime$). Note that the rotation angles were calculated counterclockwise from the original position of the features. The rotation speed of the filament's leg experienced a slow rise, then a fast increase, and finally a slow decrease. The rotation speed decreased from the inside to the outside.

Hood \& Priest (1981) used a numerical method to investigate the stability of cylindrically symmetric magnetic fields. They found that a force-free loop of uniform twist becomes unstable if its length is 3.9 times greater than its width. The loop is unstable if the twist is greater than 2.49$\pi$. For this filament, the ratio of the length (130$^\prime$$^\prime$; 94.38 Mm) to the width (31$^\prime$$^\prime$; 22.5 Mm) is about 4.2 (see Fig. 2a). The twist (2.83$\pi$) of this filament is greater than 2.49$\pi$. In fact, the twist of this filament was underestimated, because we just measured the twist from the rotation of the filament. Some twist still existed in the filament itself. Even so, the filament still satisfied the condition of kink instability.



Taking $\Phi$ as the total rotation in angle $\theta$, the twist goes from one end of the flux rope to the other end. We can get the following equation:

\begin{equation}
\label{2}
k = \frac{2 \pi}{\lambda} =\frac{\Phi}{L}, or, kR = \frac{2\pi R}{\lambda} = \frac{\Phi R}{L},
\end{equation}
where $L$, $\lambda$, and $k$ are the length of the flux rope, the helix pitch length, and the twist of unit length. We considered the filament as the cylindrically symmetric flux rope. The radius (R) is 15.5$^\prime$$^\prime$ (about 11.25 Mm) and the length (L) of the filament is 130$^\prime$$^\prime$ (94.38 Mm) (see Fig. 2a). Thus, the ratio of the radius (11.25 Mm) to the length (94.38 Mm) is 0.12. From the rotation of the leg of the filament, we obtained the rotation angle ($\Phi$) is about 2.83$\pi$. Hence, $kR$ $\approx$ 1.06 $>$ 0.654. If $kR$ is larger than 0.654, the flux rope becomes unstable (Rust \& Kumar 1996). Thus, the condition of kink instability is satisfied for this S-shaped filament.

Previous theoretical works evidenced that the occurrence of torus instability of a solar filament depends on the decay index of the background magnetic fields. The decay index is defined as n= - d log(Bt) / d log(h)(Kliem \& Torok 2006; Liu 2008; Xu et al. 2012), in which Bt is the strength of the background magnetic fields in the transverse direction and h is the radial height above the photosphere. The background magnetic fields can be computed from the observed magnetic fields over the solar surface based on a potential filed source surface (PFSS) model (Schrijver \& De Rosa 2003). We extrapolated the potential field by using the line-of-sight magnetogram at 01:33:39 UT. Figure 6 shows the magnetogram at 01:33:39 UT overplotted with the extrapolated potential field lines (red lines). The yellow line indicates the polarity inversion line (PIL). We chose twenty-one points (marked by asterisks) along the PIL from left to right to calculate the decay index. The upper panel of Fig. 7 shows the log(Bt) versus log(h) for the twenty-one points on the PIL. The height range of 46.5 - 114.4 Mm was used from the extrapolated potential fields to calculate the decay index. The stripping fields were believed to dominate in a range of about 40-100 Mm (Liu 2008; Xu et al. 2012). The lower panel of Fig. 7 shows the decay indices derived from the extrapolated potential field along the PIL from left to right. The average decay index is 1.45.

\section{Conclusion and discussion}
In this paper, we investigated the rotation motion during the filament eruption observed by SDO on 2012 May 22.
We measured the twist of the filament by analyzing the leg rotation of the filament. The rotation angle reached 510 degrees (about 2.8$\pi$) within twenty-three minutes. The maximal rotation speed reached 100 degrees per minute (about 379.9 km/s at radius 18$^\prime$$^\prime$). According to the condition of kink instability, it is indicating that the kink instability is the trigger mechanism for this solar filament eruption.

Kink instability is suggested as the triggering mechanism for solar filament eruptions (T\"{o}r\"{o}k \& Kliem 2003), which is supported by some previous observational studies (Srivastava et al. 2010; Kumar et al. 2012). However, the twist of the filament is not easy to measure. The on-disc event studied in this paper exhibited apparent rotation of the filament leg during its eruption. It is very suitable to drive the twist in the filament. According to the theoretical and simulation study, if a critical end-to-end twist of the filament is larger than 2.75, the filament will become unstable and erupt. The twist derived from the rotation of the filament leg is larger than the critical value obtained by previous authors (Hood \& Priest 1981; T\"{o}r\"{o}k \& Kliem 2003).

Torus instability is another triggering mechanism for solar filament eruptions (Kliem \& T\"{o}r\"{o}k 2006; Schrijver et al. 2008; T\"{o}r\"{o}k et al. 2010). Kliem \& T\"{o}r\"{o}k (2006) suggested that the critical decay index ranges from 1.5 to 2.0 for torus instability. The MHD simulation gave the critical value 1.53 for the occurrence of the torus instability (T\"{o}r\"{o}k \& Kliem 2005). Liu (2008) found that if the decay index is lower than 1.71, the filament will experience failed eruption. The average decay index is 1.45 for the event studied in the paper. Therefore, we think the torus instability is not the main triggering mechanism for this event.

The sunspot rotation can twist the magnetic field lines and the twist can be transferred from the photosphere to the corona (Yan et al. 2012).
Similarly, the vortex motion in the photosphere may be important for the energy and material transmission from lower atmosphere to higher atmosphere.
Wedemeyer-B\"{o}hm et al. (2012) and Yan et al. (2013) found that the energy and twist can be transmitted by vortex motion at the magnetic footpoints on small spatial
scales from the inner atmosphere to the outer solar atmosphere via magnetic fields. Except for the direct injection of twist into the erupting core flux, T\"{o}r\"{o}k et al. (2013) carried out the numerical simulation and found that the filament eruption can be triggered by sunspot rotation, which twist the potential field overlying the filament and weaken the magnetic tension over the filament. In this event, the vortex motion in the photosphere was found at the right footpoint of the filament before its eruption (see Fig. 10 of Yan et al. 2013). After the filament eruption, the leg of the filament exhibited the rapid rotation motion. This rotation of the right leg of the filament can be explained as the unwinding of the filament.

\acknowledgments
The authors thank the referees for their constructive suggestions.
SDO is a mission of NASA's Living With a Star Program. The authors are indebted to the SDO team for providing the
data. This work is supported by the National Science Foundation of
China (NSFC) under grant numbers 11373066, 11373065, 11178016, Yunnan Science Foundation of China under number
2013FB086, the Talent Project of Western Light of Chinese Academy of Sciences, the National Basic Research Program of China 973
under grant number G2011CB811400, HeBei Natural Science Foundation of China under number A2010001942, and the Foundation of Key Laboratory of Modern Astronomy and Astrophysics (Nanjing University).

\begin{figure}
\epsscale{.80}
\plotone{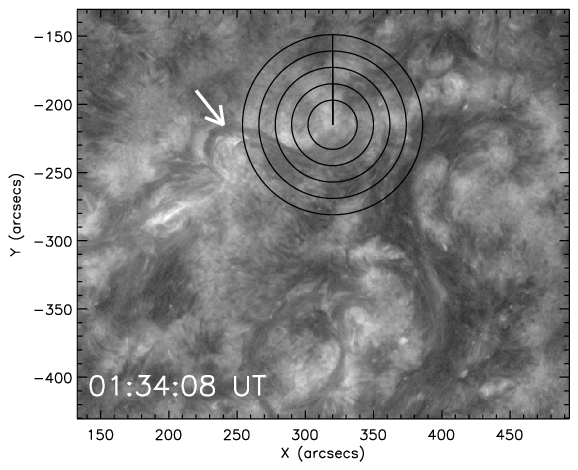}
\caption{Diagram illustrating how the leg of the filament is uncurled by using 304 \AA\ images observed by SDO. The uncurling starts at a northward pointing chord and proceeds counterclockwise around the axis of the filament. Five circles are drawn around the axis of the filaments. The difference of the radius of adjacent circles is 12$^\prime$$^\prime$ from the inside to the outside. The white arrow indicates the inverse S-shaped filament. \label{fig1}}
\end{figure}

\begin{figure}
\epsscale{.90}
\plotone{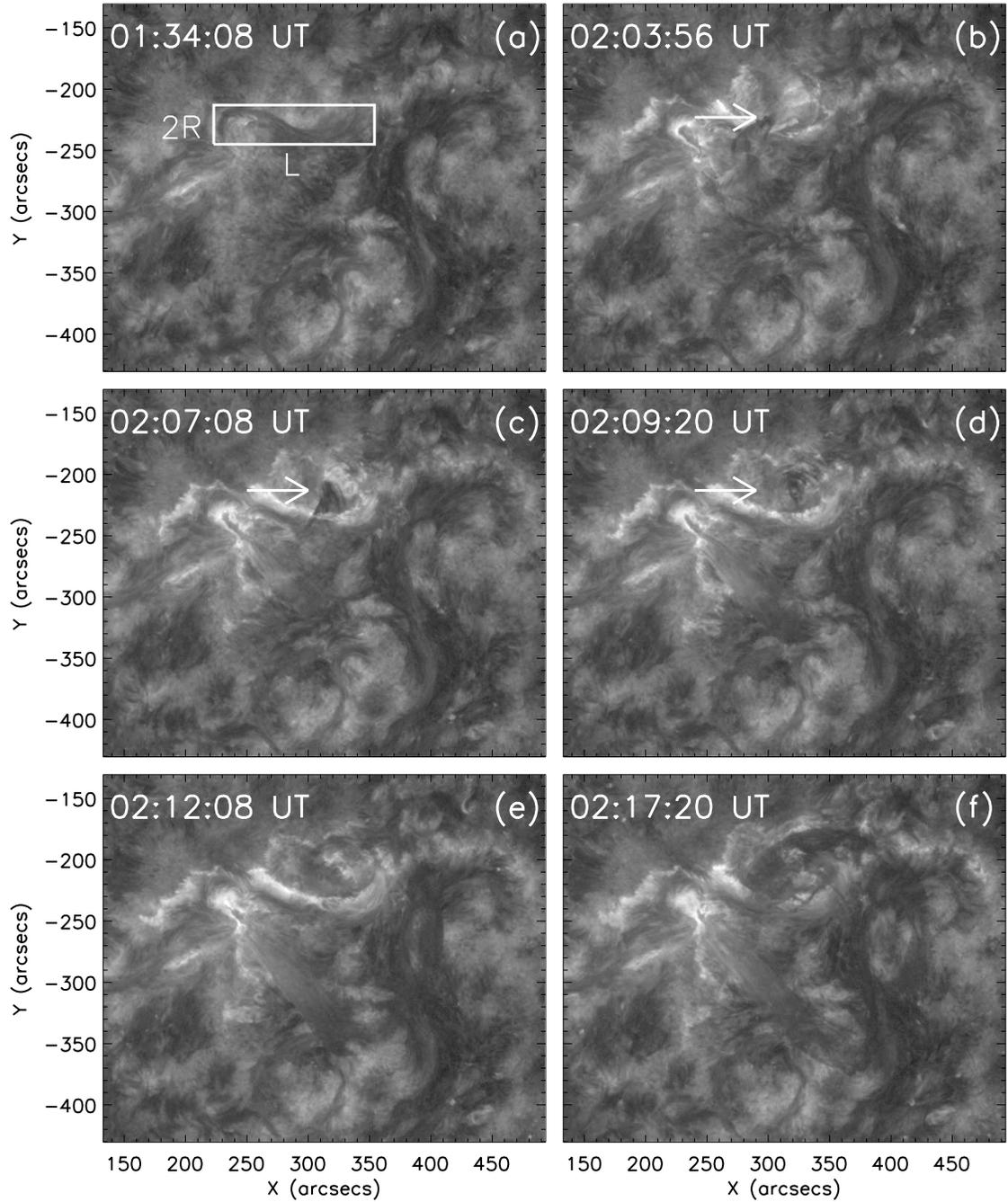}
\caption{A sequence of 304 \AA\ images to show the eruption process
of the filament from 01:34:08 UT to 02:17:20 UT on 2012 May 22. The white box in the figure denotes the position of the filament in NOAA AR 11458 and the arrows denote the leg of the filament. 2R and L denote the width and the length of the sigmoid filament, respectively.\label{fig2}}
\end{figure}

\begin{figure}
\epsscale{.90}
\plotone{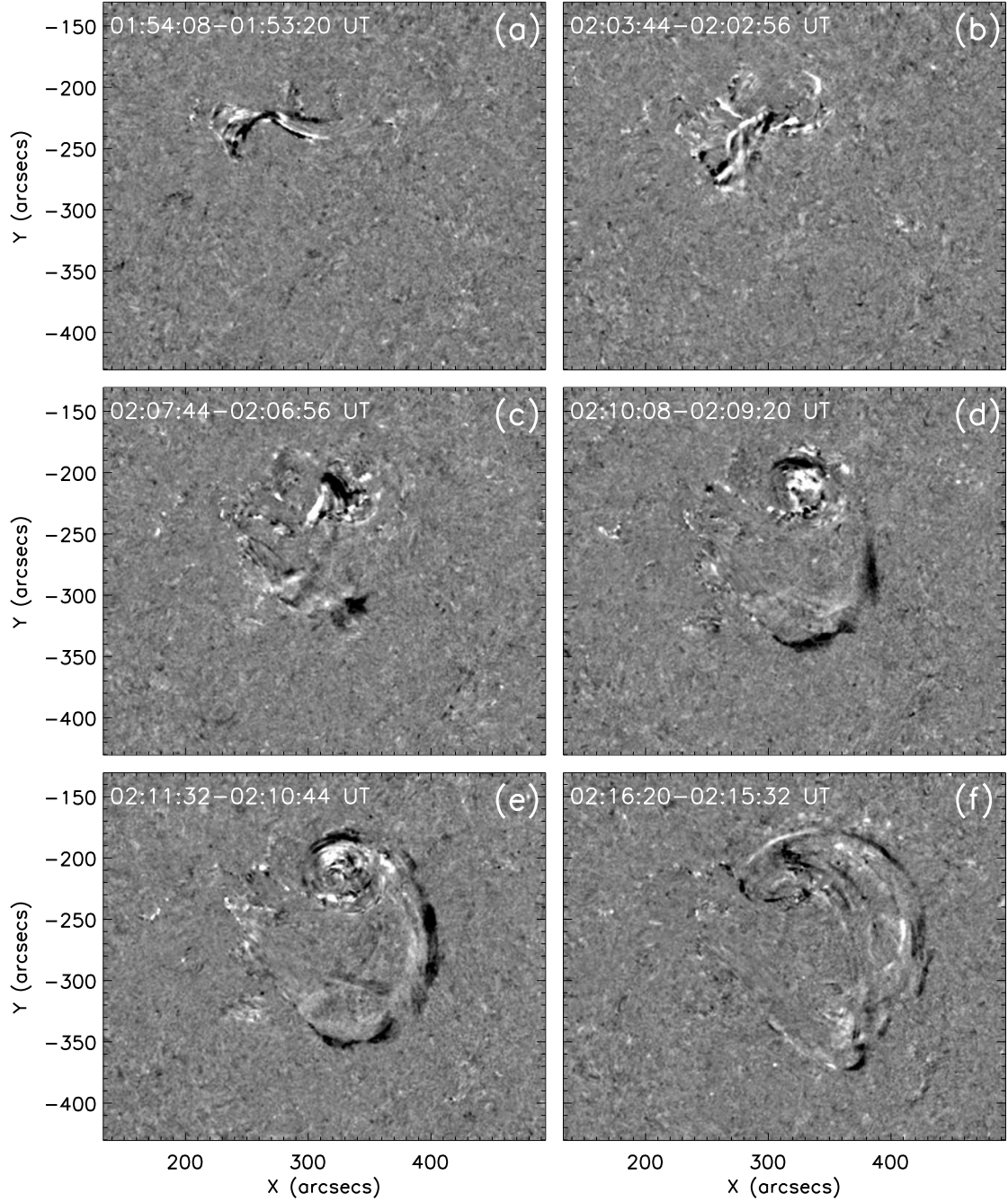}
\caption{A sequence of 304 \AA\ running-difference images to show the eruption process
of the filament from 01:54:08 UT to 02:16:20 UT on 2012 May 22. \label{fig2}}
\end{figure}

\begin{figure}
\includegraphics[angle=0,scale=1.30]{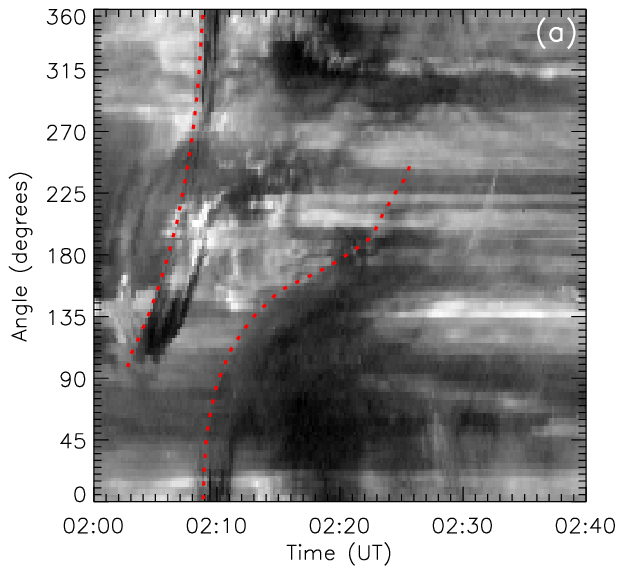}%
\includegraphics[angle=0,scale=1.30]{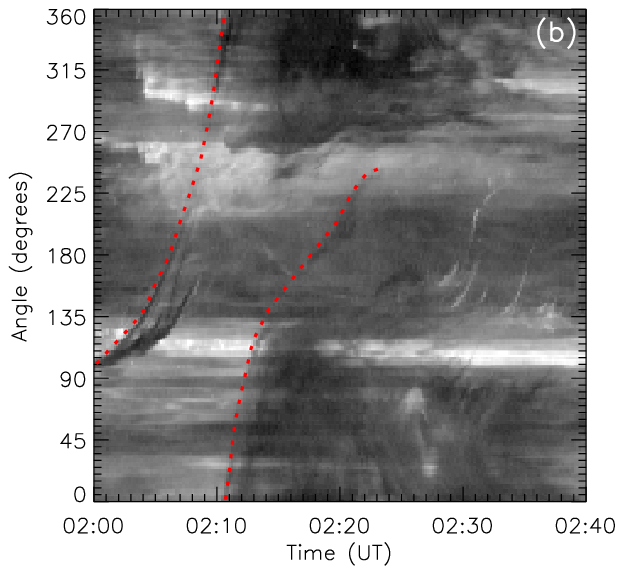}\\
\includegraphics[angle=0,scale=1.30]{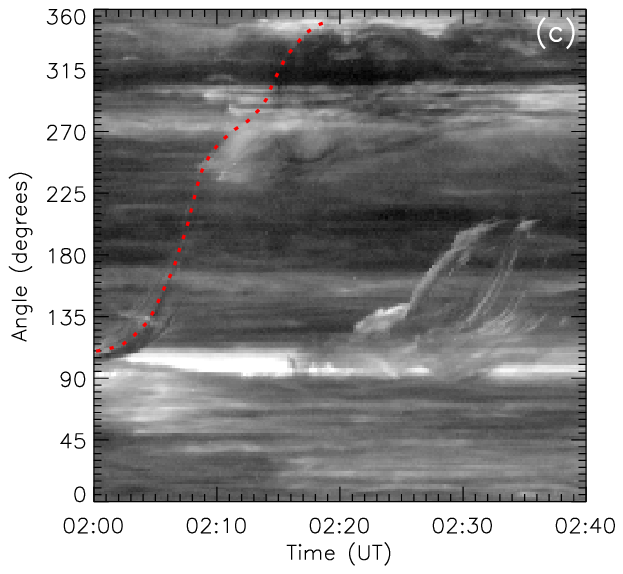}%
\includegraphics[angle=0,scale=1.30]{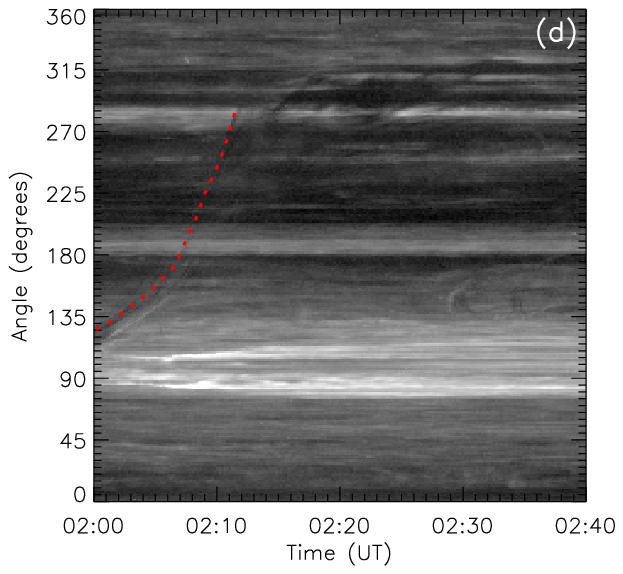}\\
\caption{Time-slices taken from the 304 \AA\ images at constant radii 18$^\prime$$^\prime$ (a), 30$^\prime$$^\prime$ (b), 42$^\prime$$^\prime$ (c), and 66$^\prime$$^\prime$ (d) from the center of the leg of the filament, respectively. Rotation can be seen from the black and white features as they move. The resolution used in the angular direction is 1$^{\circ}$. The rotation angles were calculated along the red dashed lines.}
\end{figure}

\begin{figure}
\epsscale{.80}
\plotone{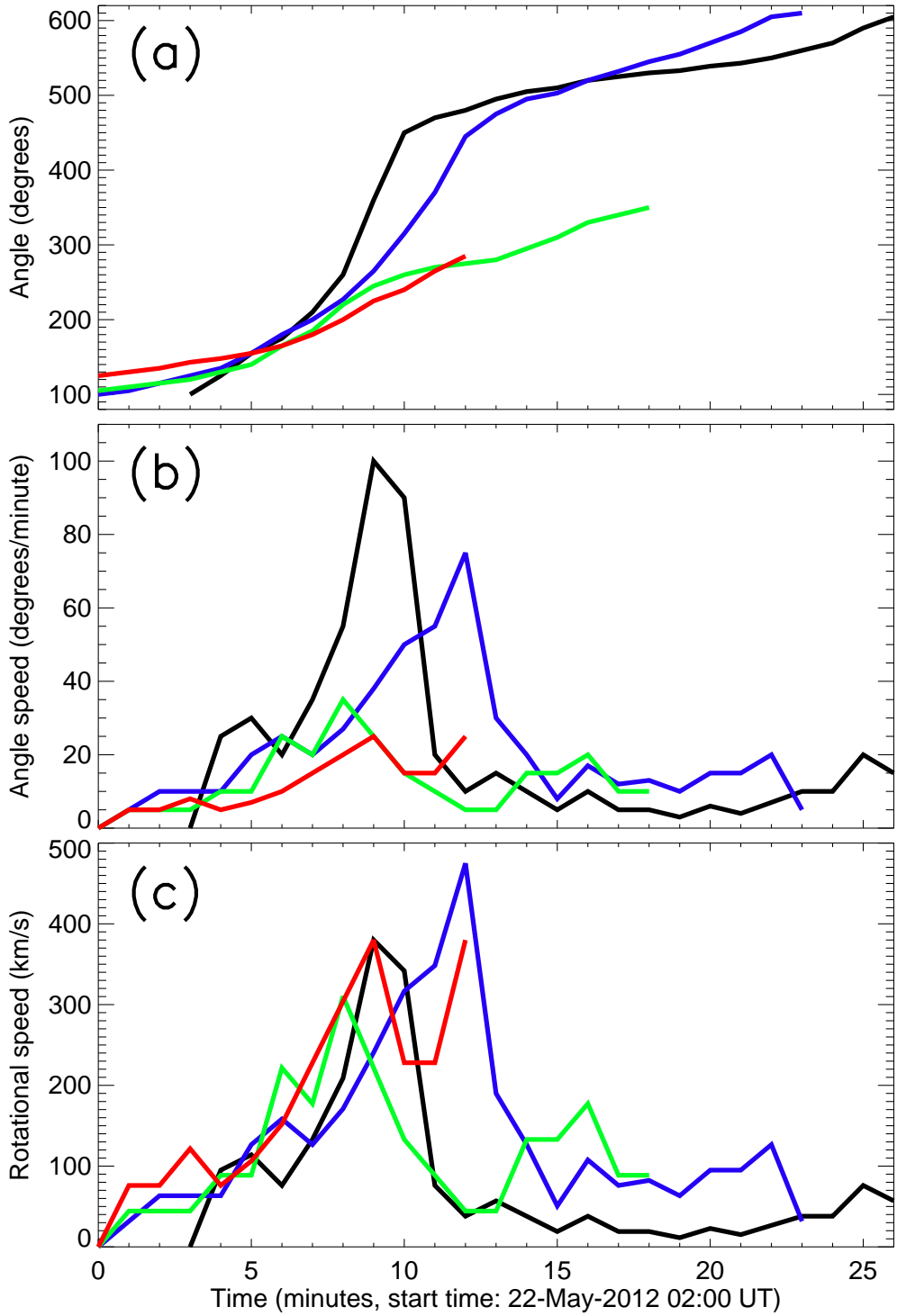}
\caption{Plot showing the rotation angle (degrees), angle speed (degree/minute), and rotation speed (km/s) at a constant radii 18$^\prime$$^\prime$ (black line), 30$^\prime$$^\prime$ (blue line), 42$^\prime$$^\prime$ (green line), and 66$^\prime$$^\prime$ (red line), respectively.\label{fig4}}
\end{figure}

\begin{figure}
\epsscale{.80}
\plotone{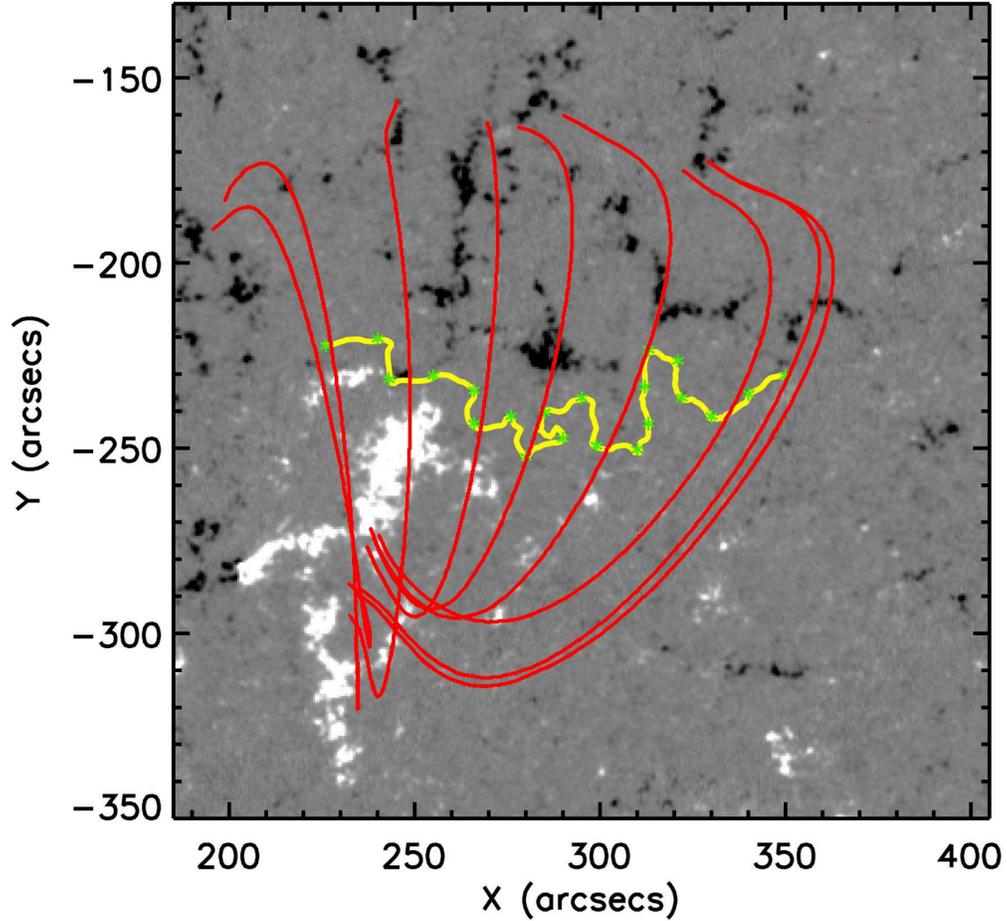}
\caption{The line-of-sight HMI magnetogram taken at 01:33:39 UT before the filament eruption, overplotted with the extrapolated potential field lines (red). The yellow line indicates the polarity inversion line (PIL). The asterisks indicate the positions used to calculate the decay index along the PIL. \label{fig5}}
\end{figure}

\begin{figure}
\epsscale{.80}
\plotone{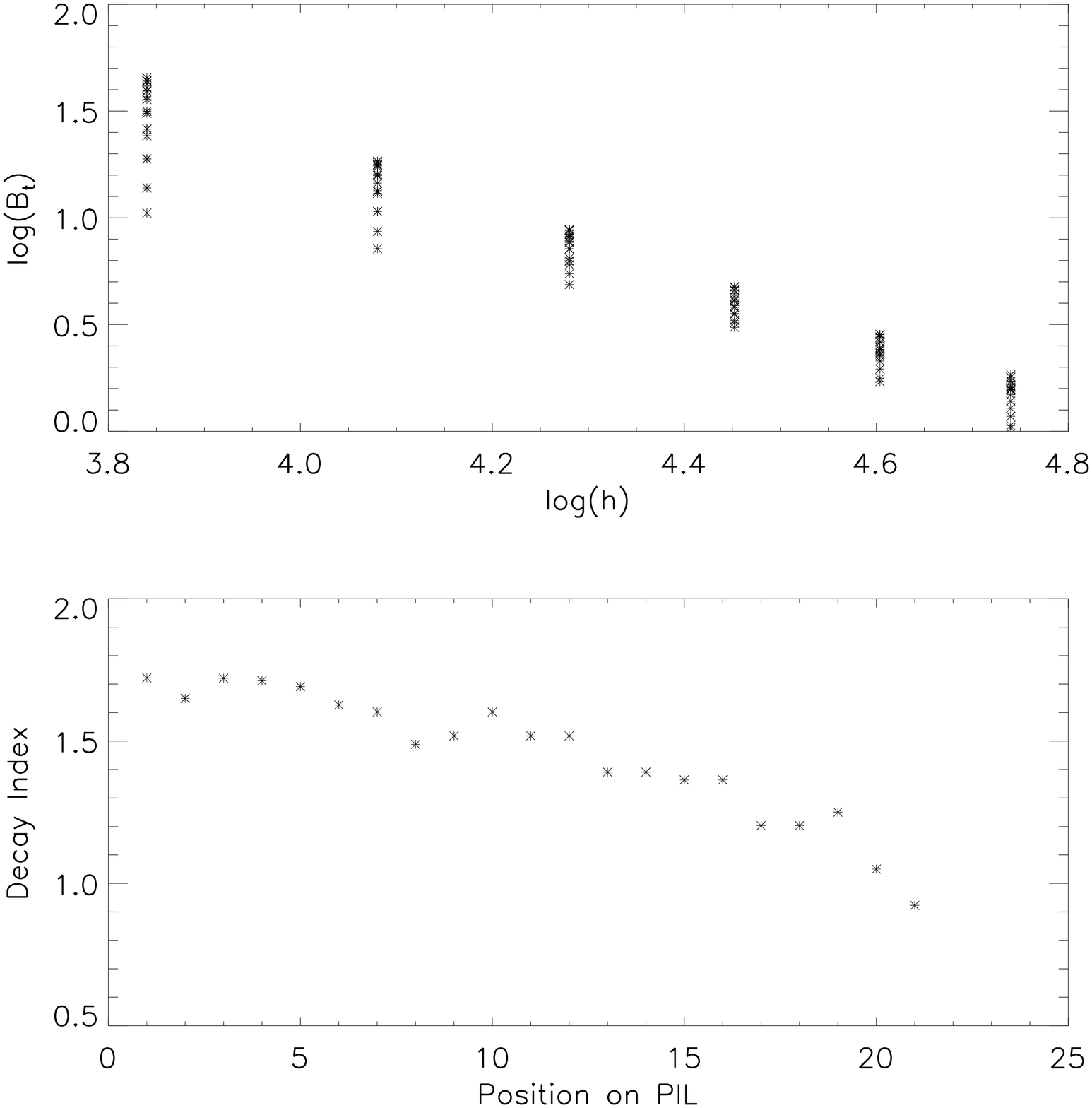}
\caption{Upper panel: The log(Bt ) vs. log(h) along the PIL. A height range of 46.5-114.5 Mm was used from extrapolated fields to calculate the decay index.  Lower panel: The decay indices derived from the extrapolated potential field along the PIL from left to right. On average, the decay index along the PIL is 1.45.\label{fig4}}
\end{figure}

\end{document}